\documentclass[aps,prl,twocolumn,groupedaddress,floatfix,nofootinbib]{revtex4-2}
\usepackage{amsmath,amssymb}
\usepackage{graphicx}
\graphicspath{{prl-figs/}}
\usepackage{xcolor}
\usepackage[colorlinks=true,linkcolor=blue!50!black,citecolor=blue!50!black,urlcolor=blue!50!black]{hyperref}

\newcommand{\E}{\mathbb{E}}
\newcommand{\R}{\mathbb{R}}
\newcommand{\la}{\langle}
\newcommand{\ra}{\rangle}
\newcommand{\I}{\mathrm{I}}
\newcommand{\Sp}{\mathrm{S}}

\begin{document}

\title{The solution of the Hopfield model with Gaussian patterns}

\author{Francesco Camilli}
\author{Pierluigi Contucci}
\author{Emanuele Mingione}
\email{emanuele.mingione2@unibo.it}
\affiliation{Department of Mathematics, Alma Mater Studiorum -- Universit\`a di Bologna, Piazza di Porta San Donato 5, 40126 Bologna, Italy}

\begin{abstract}
We prove that the free energy of the Hopfield model with Gaussian patterns has a thermodynamic limit at every load $\alpha$ and temperature $T$, and we give it as an explicit variational principle. The Hubbard--Stratonovich transformation maps the model onto $N$ Ising spins coupled to $M=\alpha N$ Gaussian hidden units, whose radius, the empirical variance of the hidden layer, is a thermodynamic variable with an entropic cost. At fixed radius the model is an Ising--spherical spin glass with Gaussian couplings, which we solve. The principle (i) reproduces the Amit--Gutfreund--Sompolinsky theory as its replica-symmetric restriction, (ii) identifies $T_g=1+\sqrt\alpha$ as the rigorous boundary of the annealed region, (iii) selects a unique spin-glass solution of the AGS equations, and (iv) gives the radius an equation of state. Moreover, (v) below $T_g$ we prove that replica symmetry is broken throughout: the instability of the AGS solution reduces to the zero-field de Almeida--Thouless condition of the Sherrington--Kirkpatrick model.
\end{abstract}

\maketitle

\paragraph{Introduction.}
The Hopfield model \cite{Hop82} is the paradigm of an associative memory: $N$ Ising spins $\sigma\in\{\pm1\}^N$ interact through $M$ stored patterns $\xi^1,\dots,\xi^M\in\R^N$ via the Hamiltonian
\begin{equation}\label{eq:H}
H_N(\sigma)=-\frac1{2N}\sum_{\mu=1}^M\big(\xi^\mu\cdot\sigma\big)^2 ,
\end{equation}
and the interesting regime is the proportional one, $M/N\to\alpha>0$, where the replica theory of Amit, Gutfreund and Sompolinsky (AGS) \cite{AGS85,AGS87} predicts a paramagnetic phase, a spin-glass phase below $T_g=1+\sqrt\alpha$, and retrieval states at small load. Rigorously, the free energy is controlled only in the replica-symmetric regime, at small $\alpha$ or high $T$ \cite{PST94,BG98,Tal98,Tal11}; for Gaussian patterns the replica-symmetric solution was derived by interpolation in \cite{BGG10} and the annealed region $T\ge T_g$ is known \cite{BG08,BGGT12}; outside these regimes not even the existence of the thermodynamic limit had been established. The reason is structural. The Hopfield Hamiltonian is not Gaussian; its Hubbard--Stratonovich representation is a bipartite spin glass between the visible spins and a layer of $M$ Gaussian hidden units, i.e.\ a restricted Boltzmann machine with continuous hidden units \cite{BGG11,BBSC12}. The covariance of this bipartite field between two replicas is the product $R_\sigma R_z$ of the overlaps of the two species, and it is not a convex function of the pair $(R_\sigma,R_z)$. The Guerra--Toninelli interpolation \cite{GT02} and the multi-species theory \cite{BCMT15,Pan15} rest on that convexity, which gives a sign to the interpolation error: for a product they do not apply.

In this Letter we derive the large $N$ limit of the  quenched pressure of the Hopfield model with Gaussian patterns, $p_N(\beta)=\frac1N\log\frac1{2^N}\sum_\sigma e^{-\beta H_N(\sigma)}$, showing it converges at every $\alpha,\beta>0$ to a variational principle in which the radius of the hidden layer appears as a thermodynamic variable. Detailed  proofs will be given in a forthcoming paper \cite{CCM26}; here we give the result, the mechanism behind it, and its consequences.

\paragraph{The result.}
Let $\mathcal N(0,s^2)$ denote the centered Gaussian of variance $s^2$ and $D(\cdot\|\cdot)$ the relative entropy. Set
\begin{equation}\label{eq:I}
I_\alpha(s)=\alpha\,D\big(\mathcal N(0,s^2)\,\big\|\,\mathcal N(0,1)\big)=\frac\alpha2\big(s^2-1-\log s^2\big).
\end{equation}
Then, almost surely and in expectation,
\begin{equation}\label{eq:main}
\lim_{N\to\infty}p_N(\beta)=P(\alpha,\beta)=\sup_{s>0}\Big\{\varphi_\alpha(\beta s^2)-I_\alpha(s)\Big\},
\end{equation}
where $\varphi_\alpha(\gamma)$ is the limit pressure of a bipartite spin glass with $N$ Ising spins and $M$ \emph{spherical} coordinates $\tau\in\sqrt M\,\mathbb S^{M-1}$, coupled by $\sqrt{\gamma/N}\sum_{i\mu}\xi_{i\mu}\sigma_i\tau_\mu$. The spherical layer appears because a Gaussian vector is a point on a sphere with a random radius $s$: at fixed radius the hidden units are spherical and $\beta s^2$ is the effective coupling. The mixed Ising--spherical pressure $\varphi_\alpha$ is itself explicit, as a variational formula over the Parisi order parameters of the two species, Eqs.~\eqref{eq:phi}--\eqref{eq:hopf} below. The supremum in \eqref{eq:main} is attained at a unique radius $s^*(\alpha,\beta)$.

The annealed pressure of the model is $P_{\rm ann}=-\frac\alpha2\log(1-\beta)$ for $\beta<1$ and $+\infty$ for $\beta\ge1$. The mixed pressure is at most its annealed value, $\varphi_\alpha(\gamma)\le\alpha\gamma/2$, with equality in the paramagnetic phase of the mixed model; when equality holds at the optimizer, \eqref{eq:main} gives back $P_{\rm ann}$, at the annealed radius $s^2=(1-\beta)^{-1}$. For $\beta\ge1$, where $P_{\rm ann}=\infty$, the glassy correction to $\varphi_\alpha$ is what keeps $P$ finite.

\paragraph{The mechanism.}
The result rests on three steps. The first is Hubbard--Stratonovich: $e^{\beta(\xi^\mu\cdot\sigma)^2/2N}=\E_z e^{\sqrt{\beta/N}\,z(\xi^\mu\cdot\sigma)}$ with $z$ standard Gaussian, so that the partition function $Z_N(\beta)=2^{-N}\sum_\sigma e^{-\beta H_N(\sigma)}$ is that of a bipartite system of $N$ Ising and $M$ Gaussian units with the Gaussian field $\sqrt{\beta/N}\sum_{i\mu}\xi_{i\mu}\sigma_iz_\mu$. The Gaussian layer is neither Ising, nor spherical, nor compact, so the recent solutions of non-convex multi-species models \cite{CIM26,CM26,CHI26} do not apply.

The second step is the observation that a Gaussian vector in $\R^M$ is a uniform point on a sphere with a random radius: $z=s\,\tau$ with $\tau$ uniform on $\sqrt M\,\mathbb S^{M-1}$ and $s=|z|/\sqrt M$ independent, with density $\nu_M(ds)\propto s^{M-1}e^{-Ms^2/2}$. Since the field is linear in $z$, the radius factors out of the coupling, and exactly
\begin{equation}\label{eq:polar}
Z_N(\beta)=\int_0^\infty Z^{\rm mix}_N(\beta s^2)\,\nu_M(ds),
\end{equation}
where $Z^{\rm mix}_N(\gamma)$ is the partition function of the Ising--spherical model at coupling $\gamma$. The radius is the only degree of freedom that the Gaussian layer adds to the spherical one, and its distribution concentrates exponentially: $\frac1N\log\nu_M(ds)/ds=-I_\alpha(s)+o(1)$ by Stirling's formula, which explains \eqref{eq:I}. The integral in \eqref{eq:polar} is then evaluated by a Laplace principle, and \eqref{eq:main} follows, provided that $\frac1N\log Z^{\rm mix}_N(\beta s^2)$ converges and concentrates uniformly in $s$ on compacts, which it does: the field is linear in $s$, so $\log Z^{\rm mix}_N$ is convex in $s$, and pointwise convergence of convex functions is uniform on compacts. This is the equivalence-of-ensembles argument of \cite{GT15,GT20}, and the continuous counterpart of the slices of fixed magnetization recently used for Ising systems \cite{CHI26,BM26}. The third step, the solution of the mixed model, is given after the physical reading that the first two allow.

\begin{figure}[t]
\includegraphics[width=\columnwidth]{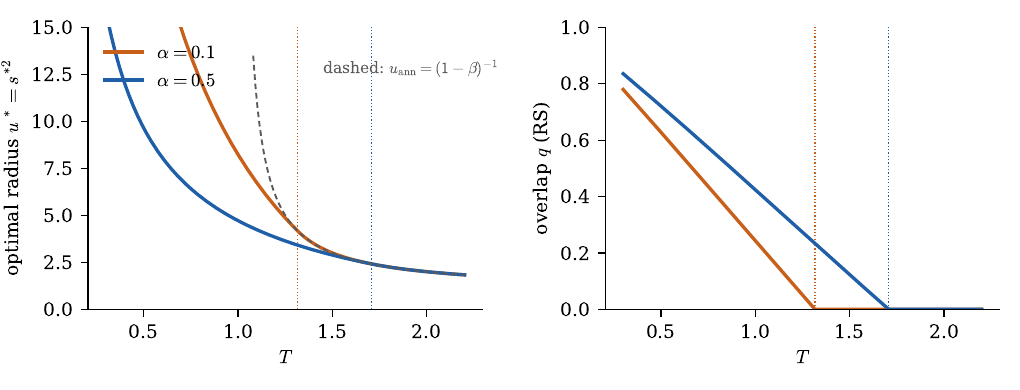}
\caption{Left: optimal radius $u^*=s^{*2}$ as a function of $T$ within the replica-symmetric class, for $\alpha=0.1$ and $\alpha=0.5$; the dashed curve is the annealed radius $(1-\beta)^{-1}$, which diverges at $T=1$. The radius leaves the annealed branch at $T_g$ (dotted) and stays finite at every temperature. Right: the corresponding Edwards--Anderson overlap $q$, which vanishes continuously at $T_g$.}
\label{fig:radius}
\end{figure}

\paragraph{The radius as a thermodynamic variable.}
Equation \eqref{eq:main} invites a physical reading: the quantity $s^2$ is the empirical variance of the hidden units, i.e.\ the amount of fluctuation that the hidden layer retains. The prior wants $s^2=1$, at entropic cost $I_\alpha(s)$ otherwise; the interaction wants a large radius, since the coupling grows with $s$; \eqref{eq:main} settles the competition. In the annealed phase the radius is dictated by energy alone, $s^2=(1-\beta)^{-1}$, which diverges at $\beta=1$: the hidden layer would absorb unbounded fluctuations. In the glassy phase the correction to the annealed value of $\varphi_\alpha$ grows with $s$ and pulls the radius back, so that $P$ stays finite at every temperature. The optimal radius is unique: by \eqref{eq:phi}--\eqref{eq:hopf}, $u\mapsto\varphi_\alpha(\beta u)$ is linear in $u=s^2$ minus $(1+\alpha)f$, with $f$ convex as a supremum of affine functions of $t$, so the functional in \eqref{eq:main} is strictly concave in $u$. Differentiating \eqref{eq:main} in $u$ gives the equation of state
\begin{equation}\label{eq:radius}
u^*=\Big(1-\beta+\frac{(1+\alpha)^2\beta}{\alpha}\int_0^1p^\I p^\Sp\,du\Big)^{-1},
\end{equation}
where $\int_0^1p^\I p^\Sp$ is the overlap term of the formula \eqref{eq:hopf} below, evaluated at the optimal Parisi order parameters. It vanishes in the paramagnet and is the glassy correction to the annealed radius: the more correlated the replicas, the smaller the fluctuation that the hidden layer keeps. Figure~\ref{fig:radius} shows $u^*(T)$ within the replica-symmetric class.

\paragraph{The mixed model.}
The third step is the solution of the Ising--spherical model, which we obtain by the Hamilton--Jacobi approach of Mourrat \cite{MouAHL,Mou21,Mou23}; the bipartite case is the one for which that approach was conjectured to give the limit \cite{Mou21,MouICM}. The idea is to regard the free energy as a function of a fictitious time $t$, which plays the role of the coupling strength (here $t=(1+\alpha)\beta s^2/2$), and of a Parisi order parameter $\pi=(\pi^\I,\pi^\Sp)$: a pair of nondecreasing functions on $[0,1)$ encoding the overlap distribution of the two species, which share one ultrametric tree. One shows that this function satisfies a first-order Hamilton--Jacobi equation whose Hamiltonian is the covariance of the model, here the product $\kappa(a,b)=ab$ of the two overlaps, and whose initial condition at $t=0$ is the entropic part of the Parisi functional, $\psi(\pi)=\lambda_\I\psi_\I(\pi^\I)+\lambda_\Sp\psi_\Sp(\pi^\Sp)$, with $\lambda_\I=\frac1{1+\alpha}$ and $\lambda_\Sp=\frac\alpha{1+\alpha}$ the fractions of the two species: $\psi_\I$ is the free energy of one Ising spin in the random field with overlap structure $\pi^\I$, the nested $\log\cosh$ of the Parisi recursion, and $\psi_\Sp$ the Crisanti--Sommers expression for one spherical coordinate \cite{CIM26,CM26}. When the Hamiltonian $\kappa$ is convex, as for the SK model, the solution is a variational principle in the variable $\pi$ itself: the Parisi formula. When $\kappa$ is not convex but the initial condition $\psi$ is, as here, the solution is the Hopf formula: a variational principle in the conjugate variable $p=\partial_\pi\psi(\pi)$, in which $\kappa$ enters linearly and the Legendre transform $\psi^*(p)=\sup_\pi\{\int_0^1 p\,\pi-\psi(\pi)\}$ takes the place of $\psi$. Physically, $\pi$ is the Parisi function seen by the entropic term, $p$ the overlap seen by the energy (in the replica-symmetric case $p^\I=\lambda_\I q$, with $q$ the Edwards--Anderson overlap); the Parisi formula identifies them at the stationary point, the Hopf formula keeps them conjugate, and for SK the two agree. In the present case,
\begin{equation}\label{eq:phi}
\varphi_\alpha(\gamma)=\frac{\alpha\gamma}2-(1+\alpha)\,f\Big(\frac{(1+\alpha)\gamma}2\Big),
\end{equation}
where $1+\alpha=(N+M)/N$ converts the pressure per unit of the mixed model into that per visible spin, and
\begin{equation}\label{eq:hopf}
f(t)=\sup_{p=(p^\I,p^\Sp)}\Big\{t\int_0^1p^\I(u)\,p^\Sp(u)\,du-\psi^*(p)\Big\},
\end{equation}
the supremum running over nondecreasing $p^\I\le\lambda_\I$, $p^\Sp\le\lambda_\Sp$. Equation \eqref{eq:main} is thus a principle of the sup--inf type: a supremum over the radius and over the overlap paths $p$, an infimum over the Parisi order parameters $\pi$, the $-\sup_\pi$ inside $-\psi^*$. This is the form the Hamilton--Jacobi approach assigns to all non-convex spin glasses \cite{Mou21,CM25,CIM26,CM26}.

The Hamilton--Jacobi equation gives directly the lower bound; the upper bound is the delicate part, and its logic is the rigorous content of the replica method for non-convex interactions. A cavity computation, under the Ghirlanda--Guerra perturbations that enforce ultrametricity, shows that the free energy is a stationary value of the functional in \eqref{eq:hopf}, at a pair $p$ with $p=\partial_\pi\psi(\pi)$ at $\pi=(tp^\Sp,tp^\I)$, the Parisi variable of each species being the effective field of the other \cite{CM25}; the method does not say which stationary point. What selects it is the convexity of $\psi$ in $\pi$: every stationary point is then a minimizer, and the upper bound coincides with the Hopf formula. Convexity holds for centered Ising spins \cite{CIM26} and for spherical spins \cite{CM26}, hence for their sum $\psi$: this is the structural fact behind \eqref{eq:hopf}. The cavity computation for the mixed measure is the technical part of \cite{CCM26}: the block cavity of \cite{CM26} extends to it because Ising spins have fixed self-overlap.

\begin{figure}[t]
\includegraphics[width=\columnwidth]{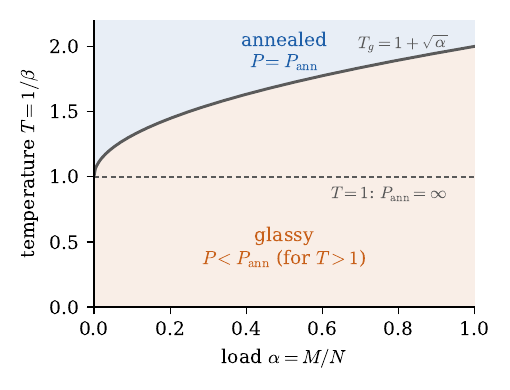}
\caption{Phase diagram of the Hopfield model with Gaussian patterns in the $(\alpha,T)$ plane. Above $T_g=1+\sqrt\alpha$ the pressure is annealed; below, it is strictly smaller than the annealed one, which diverges at $T=1$, and replica symmetry is broken. There is no retrieval phase at positive load.}
\label{fig:phase}
\end{figure}

\paragraph{Phase diagram.}
The variational principle gives the phase diagram of Fig.~\ref{fig:phase}. For $\beta<1$, $P\le P_{\rm ann}$ with equality if and only if $\beta\le\beta_g=1/(1+\sqrt\alpha)$: above $T_g=1+\sqrt\alpha$ the pressure is annealed, recovering the theorem of \cite{BGGT12}, and below $T_g$ it is strictly smaller, which is new as a theorem. The threshold is produced by the formula itself. The paramagnetic phase of the mixed model is the region where $f(t)=0$, i.e.\ where the pair $p=0$ is optimal in \eqref{eq:hopf}. Near $p=0$ the entropic functional is quadratic, $\log\cosh x\simeq x^2/2$, and in fact $\psi(\pi)\le\int(\lambda_\I(\pi^\I)^2+\lambda_\Sp(\pi^\Sp)^2)$ at every path, a bound whose Legendre transform is $\psi^*(p)\ge\int\big(\frac{(p^\I)^2}{4\lambda_\I}+\frac{(p^\Sp)^2}{4\lambda_\Sp}\big)$. The functional in \eqref{eq:hopf} is therefore at most $\int\big(t\,p^\I p^\Sp-\frac{(p^\I)^2}{4\lambda_\I}-\frac{(p^\Sp)^2}{4\lambda_\Sp}\big)$, a quadratic form in $(p^\I,p^\Sp)$ at each $u$, negative definite if and only if $t<t_c=\frac1{2\sqrt{\lambda_\I\lambda_\Sp}}=\frac{1+\alpha}{2\sqrt\alpha}$; since also $f\ge0$ (take $p=0$), $f(t)=0$ for $t\le t_c$. At the annealed radius $t=\frac{(1+\alpha)\beta}{2(1-\beta)}$, and $t\le t_c$ reads $\beta\le\beta_g$. Conversely, for $t>t_c$ the small constant pair $p=\varepsilon(\sqrt{\lambda_\I},\sqrt{\lambda_\Sp})$ gives $f(t)\ge\varepsilon^2\sqrt{\lambda_\I\lambda_\Sp}\,(t-t_c)+O(\varepsilon^3)>0$, so that $P<P_{\rm ann}$ for $\beta>\beta_g$. This is the linear instability of the paramagnetic AGS solution, where a nonzero solution of the linearized equations $q\simeq\alpha\beta^2r$, $r\simeq q/(1-\beta)^2$ bifurcates at $\sqrt\alpha\,\beta=1-\beta$; here it is read off an exact formula, as the case $\kappa_\I=\kappa_\Sp=2$ of the replicon condition \eqref{eq:replicon}. Since $P_{\rm ann}$ is analytic, $P$ is not analytic at $\beta_g$.

For Gaussian patterns the replica theory predicts no retrieval phase at positive load \cite{BGST18}, and a symmetry argument makes this a theorem for the symmetric Gibbs measure: $\xi O$ has the law of $\xi$ for every orthogonal $O$ and $H_N$ is invariant, so the vector of Mattis overlaps $m_\mu=\xi^\mu\cdot\sigma/N$ is, conditionally on its norm, uniform on a sphere of $\R^M$, and $\E\la\max_\mu m_\mu^2\ra\le C\log M/M$ at every $(\alpha,\beta)$. Retrieval is a $\pm1$-pattern phenomenon.

\paragraph{Replica-symmetric restriction and AGS.}
Restricting the supremum in \eqref{eq:hopf} to constant paths, $p=(\lambda_\I x,\lambda_\Sp y)$ with $x$ the Ising and $y$ the spherical overlap, gives a replica-symmetric pressure $P_{\rm RS}\ge P$. The Legendre transform $\psi^*$ then reduces to the transforms of the one-species functions $\Psi_\I(c)=c-\E\log\cosh(\sqrt{2c}\,z)$ and $\Psi_\Sp(c)=\frac{(\bar b-1)(\bar b-2)}2+\frac12\log\bar b$, $\bar b=\frac{1+\sqrt{1+8c}}2$, $c$ being the variance of the replica-symmetric cavity field. The supremum over the spherical overlap $y$ and over the radius can be carried out explicitly, and what remains is
\begin{equation}\label{eq:AGS}
P_{\rm RS}=\inf_{q}\sup_{r\ge0}\mathcal A_{\alpha,\beta}(q,r),
\end{equation}
where $\mathcal A_{\alpha,\beta}(q,r)=-\frac\alpha2[\log(1-\beta+\beta q)-\frac{\beta q}{1-\beta+\beta q}]-\frac{\alpha\beta^2}2r(1-q)+\E\log\cosh(\beta\sqrt{\alpha r}z)$ is exactly the AGS replica-symmetric functional at zero Mattis magnetization \cite{AGS87,BGG10}, in our normalization, $q=x$ being the Edwards--Anderson overlap of the visible spins and $r$ its conjugate; the stationarity conditions of \eqref{eq:AGS} are the AGS equations $q=\E\tanh^2(\beta\sqrt{\alpha r}z)$, $r=q/(1-\beta+\beta q)^2$. The variational principle also fixes the prescription, which the replica method leaves ambiguous. The replica-symmetric pressure is strictly concave in the radius, and its inner supremum over $x$ has at most one critical point, by the Lata\l a--Guerra argument \cite{Tal11}: hence the supremum over the radius and the infimum over $q$ can be exchanged, which turns the inequality $P_{\rm RS}\le\inf_q\sup_r\mathcal A_{\alpha,\beta}$ into the identity \eqref{eq:AGS}; moreover the AGS equations have a spin-glass solution $q>0$ if and only if $\beta>\beta_g$, and it is unique (End Matter). The AGS free energy of the spin-glass solution is therefore a rigorous upper bound on the pressure: this is the first assumption-free derivation of the AGS functional as an upper bound at all temperatures. The identity \eqref{eq:AGS} was also checked numerically.

\begin{figure}[t]
\includegraphics[width=\columnwidth]{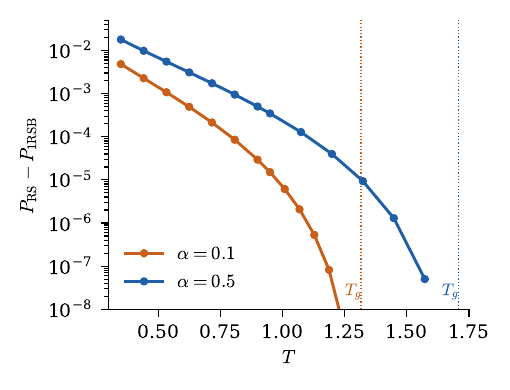}
\caption{Gain $P_{\rm RS}-P_{\rm 1RSB}$ of the one-step bound over the replica-symmetric one as a function of $T$, for $\alpha=0.1$ and $\alpha=0.5$ (log scale). It is strictly positive below $T_g$ (dotted) by \eqref{eq:strict} and vanishes continuously at $T_g$; points are shown down to the numerical resolution, $10^{-8}$.}
\label{fig:gain}
\end{figure}

\paragraph{Replica symmetry breaking.}
Below $T_g$ the replica-symmetric bound is never sharp. Any trial pair of paths in \eqref{eq:hopf} gives an upper bound on $P$: constant paths give $P_{\rm RS}$; one-step paths, with two values on $[0,\zeta)$ and $[\zeta,1)$, give $P_{\rm 1RSB}\le P_{\rm RS}$. The replica-symmetric path is a critical point of the Hopf functional, and its second variation along a zero-mean one-step perturbation, the replicon mode, is positive if and only if
\begin{equation}\label{eq:replicon}
t^2\lambda_\I\lambda_\Sp\,\kappa_\I\kappa_\Sp>1,
\end{equation}
where $\kappa_\I=2\E\,{\rm sech}^4(\sqrt{2c}\,z)$ and $\kappa_\Sp=2/\bar b^2$ are the replicon eigenvalues of the two single-species functionals at the cavity-field variance $c$ (End Matter). At the spin-glass solution the two species drop out of the condition, which becomes
\begin{equation}\label{eq:AT}
\sigma^2\,\E\,{\rm sech}^4(\sigma z)>\E\tanh^2(\sigma z),\qquad \sigma^2=2c=\alpha\beta^2r,
\end{equation}
precisely the zero-field de Almeida--Thouless instability of the SK model \cite{AT78}, as also found in \cite{AGS87}. Inequality \eqref{eq:AT} holds for every $\sigma>0$, by the sum of squares \eqref{eq:SOS}, the strict Gaussian Poincar\'e inequality for $\tanh(\sigma z)$. Hence, for every $\alpha>0$ and $\beta>\beta_g$,
\begin{equation}\label{eq:strict}
P\le P_{\rm 1RSB}<P_{\rm RS}:
\end{equation}
the optimal paths of \eqref{eq:hopf} are never constant below $T_g$, the whole spin-glass phase is replica-symmetry-broken, and the AGS free energy is a strict upper bound there. This is the analogue of Toninelli's result for SK beyond the AT line \cite{Ton02}, with the Hopf bound in place of Guerra's \cite{Gue03}, which is not available for a bilinear covariance. Figure~\ref{fig:gain} illustrates the gain $P_{\rm RS}-P_{\rm 1RSB}$, which vanishes continuously at $T_g$; the optimal radius decreases with respect to its replica-symmetric value, as the equation of state \eqref{eq:radius} predicts for a larger overlap term.

\paragraph{Discussion.}
The principle \eqref{eq:main} settles the existence of the limit, recovers the AGS theory as a rigorous upper bound, strict throughout the spin-glass phase, and completes it with a new thermodynamic variable, the radius of the hidden layer. It is useful to emphasize that the replica computation integrates the Gaussian hidden layer exactly at its saddle point: the radius never appears in it, nor does the sup--inf structure in which the Parisi variable and the overlap remain conjugate rather than identified. Open questions: whether the breaking is full as expected for SK at zero external field, and the structure of the two-species Parisi measure; whether the supremum over the radius and the infimum hidden in $\psi^*$ can be exchanged; and whether a field on the visible spins, which breaks the convexity of the Ising initial condition \cite{CIM26}, can be handled by the magnetization constraint of \cite{BM26} inside the mixed cavity. The $\pm1$ case, where retrieval lives, requires a treatment of condensed overlaps beyond the present method. The radial decomposition applies to any Gaussian layer, one Laplace principle per layer: deep Boltzmann machines with Gaussian layers are the natural next step.

\begin{acknowledgments}
\textbf{AI use statement} The proof strategy is an original idea of the authors. For intermediate derivations, some proof details and numerical simulations, the authors were assisted by Fable 5.1 (Anthropic) and GPT-6 Astra (Open AI). The authors independently checked the resulting arguments, calculations, numerical results, and take full responsibility for the content of the present work.
\end{acknowledgments}

\appendix

\section{Uniqueness of the spin-glass solution}\label{app:LG}
We prove the replica-symmetric formula \eqref{eq:AGS} and the
existence and uniqueness of its physical spin-glass solution. Throughout,
$\alpha,\beta>0$, $u=s^2$, and $t_u=(1+\alpha)\beta u/2$.
Set $D(x)=1-\beta+\beta x$. The admissible AGS domain is
\begin{align}
\mathcal D_\beta=\{x\in[0,1):D(x)>0\}.
\label{eq:appA-domain}
\end{align}

\emph{Reduction to two scalar variables.}
The single-species conjugates are
$\Psi_s^*(x)=\sup_{c\ge0}\{cx-\Psi_s(c)\}$.
For constant overlap paths, optimization over the spherical overlap uses
\begin{align*}
\sup_{0\le y<1}\{hy-\Psi_\Sp^*(y)\}=\Psi_\Sp(h),
\end{align*}
with $h=t_u\lambda_\I x=\beta ux/2$.
Since $(1+\alpha)\lambda_\Sp=\alpha$, this yields
\begin{align}
P_{\rm RS}&=\sup_{u>0}\Lambda(u),\nonumber\\
\Lambda(u)&=\min_{0\le x<1}K(u,x),
\label{eq:Lambda}\\
K(u,x)&=A(u)-F_u(x),\nonumber\\
A(u)&=\frac\alpha2\big[(\beta-1)u+1+\log u\big],\nonumber\\
F_u(x)&=\alpha\Psi_\Sp(\beta ux/2)-\Psi_\I^*(x).
\nonumber
\end{align}
We first solve the inner problem, then optimize the radius.

\emph{(a) The inner maximizer.}
Set $g(v)=\E\tanh^2(\sqrt v\,z)$, with $z$ standard Gaussian.
This is a strictly increasing continuous bijection from $[0,\infty)$
onto $[0,1)$. Gaussian integration by parts and scalar duality give
\begin{align*}
\Psi_\I'(c)=g(2c),\qquad
(\Psi_\I^*)'(x)=\tfrac12g^{-1}(x).
\end{align*}
Also, writing
\begin{align*}
b_u(x)&=\frac{1+\sqrt{1+4\beta ux}}2,\\
Y_u(x)&=\alpha\beta u\big(1-b_u(x)^{-1}\big),
\end{align*}
we have $\Psi_\Sp'(\beta ux/2)=1-b_u(x)^{-1}$ and
\begin{align}
2F_u'(x)=Y_u(x)-g^{-1}(x).
\label{eq:appA-inner-derivative}
\end{align}
For $x>0$, its sign is therefore the sign of
$g(Y_u(x))/x-1$. The ratio $g(v)/v$ is strictly decreasing:
indeed,
\begin{align*}
\frac{g(v)}v
=\E\left[z^2\left(
\frac{\tanh(\sqrt v\,z)}{\sqrt v\,z}\right)^2\right],
\end{align*}
and $\tanh a/a$ is strictly decreasing for $a>0$.
Moreover, $Y_u$ is strictly increasing and
\begin{align*}
\frac{Y_u(x)}x=\frac{\alpha\beta^2u^2}{b_u(x)^2}
\end{align*}
is strictly decreasing. Factoring the ratio as
\begin{align*}
\frac{g(Y_u(x))}x
=\frac{g(Y_u(x))}{Y_u(x)}\frac{Y_u(x)}x
\end{align*}
shows that it is strictly decreasing, with limits
\begin{align*}
\lim_{x\downarrow0}\frac{g(Y_u(x))}x&=\alpha\beta^2u^2,
&\lim_{x\uparrow1}\frac{g(Y_u(x))}x&<1.
\end{align*}
It follows that $F_u$ has a unique global maximizer $x_u$:
\begin{align}
x_u=0\quad&(u\le u_c),&
x_u\in(0,1)\quad&(u>u_c),\nonumber\\
u_c&=\frac1{\beta\sqrt\alpha}.
\label{eq:appA-inner-threshold}
\end{align}
For $u\le u_c$, $F_u$ is strictly decreasing on $(0,1)$;
for $u>u_c$, its derivative changes sign once, from positive to
negative. In the latter case, $x_u=g(Y_u(x_u))$.

\emph{(b) The optimal radius.}
Since $\Psi_\Sp$ is convex, $u\mapsto\sup_xF_u(x)$ is convex,
whereas $A''(u)=-\alpha/(2u^2)<0$. Hence $\Lambda$ is strictly
concave. It also tends to $-\infty$ at both ends of $(0,\infty)$.
At $0$, use $\Lambda\le A$. At infinity, fix any
$x_0\in\mathcal D_\beta\cap(0,1)$ and use
$\Psi_\Sp(c)=c+O(\sqrt c)$ to obtain
\begin{align*}
\Lambda(u)\le K(u,x_0)
=-\tfrac\alpha2D(x_0)u+O(\sqrt u+\log u).
\end{align*}
Thus a unique maximizing radius $\hat u$ exists.

The inner maximizer $x_u$ is continuous in $u$, including at $u_c$.
To justify the envelope theorem despite the open endpoint $x=1$,
note that $\Psi_\I^*(x)\to\infty$ as $x\uparrow1$; hence the
maximizers stay away from $1$ locally uniformly in $u$.
Continuity then follows from uniqueness. The envelope theorem yields
\begin{align}
\Lambda'(u)=\frac\alpha2\left[
\beta-1+\frac1u
-\beta x_u\big(1-b_u(x_u)^{-1}\big)\right].
\label{eq:appA-envelope}
\end{align}
In particular, $\Lambda$ is $C^1$ and
\begin{align*}
\Lambda'(u_c)
=\frac\alpha2\big[\beta(1+\sqrt\alpha)-1\big].
\end{align*}
For $\beta\le\beta_g=(1+\sqrt\alpha)^{-1}$, the annealed radius
$u_{\rm ann}=(1-\beta)^{-1}$ satisfies $u_{\rm ann}\le u_c$ and
$\Lambda'(u_{\rm ann})=0$. Therefore $\hat u=u_{\rm ann}$ and
$x_{\hat u}=0$. For $\beta>\beta_g$, $\Lambda'(u_c)>0$, so
strict concavity implies $\hat u>u_c$ and $x_{\hat u}>0$.

\emph{(c) Optimization at fixed overlap.}
Write $K(u,x)=g_x(u)+\Psi_\I^*(x)$, where
$g_x(u)=A(u)-\alpha\Psi_\Sp(\beta ux/2)$.
This function is strictly concave in $u$, and the identity
$b_u(x)(b_u(x)-1)=\beta ux$ gives
\begin{align*}
g_x'(u)=\frac\alpha2\left[\frac{b_u(x)}u-D(x)\right].
\end{align*}
If $D(x)\le0$, then $g_x'(u)\ge\alpha/(2u)$, so
$\sup_ug_x(u)=+\infty$. If $D(x)>0$, its unique stationary
point satisfies $b_u(x)=D(x)u$. Solving for $u$ and substituting
back gives
\begin{align}
u^*(x)&=\frac{D(x)+\beta x}{D(x)^2},\nonumber\\
\sup_{u>0}g_x(u)
&=-\frac\alpha2\left[\log D(x)-\frac{\beta x}{D(x)}\right].
\label{eq:appA-fixed-overlap}
\end{align}
Define $\Phi(x)=\sup_uK(u,x)$ on $\mathcal D_\beta$, and
$\Phi(x)=+\infty$ otherwise.

\emph{(d) The saddle point and the AGS equations.}
Let $\hat x=x_{\hat u}$. By definition, $\hat x$ minimizes
$K(\hat u,\cdot)$. Equation~\eqref{eq:appA-envelope} gives
$\partial_uK(\hat u,\hat x)=0$; strict concavity in $u$ therefore
makes $\hat u$ the global maximizer of $K(\cdot,\hat x)$.
Thus $\hat x\in\mathcal D_\beta$, $\hat u=u^*(\hat x)$, and
\begin{align*}
K(u,\hat x)\le K(\hat u,\hat x)\le K(\hat u,x)
\end{align*}
for every $u>0$ and $x\in[0,1)$. This proves the minimax equality
\begin{align}
P_{\rm RS}
&=\sup_{u>0}\inf_{0\le x<1}K(u,x)\nonumber\\
&=\inf_{x\in\mathcal D_\beta}\sup_{u>0}K(u,x)
=\Phi(\hat x).
\label{eq:appA-minimax}
\end{align}
The minimizer of $\Phi$ is unique: equality $\Phi(x)=P_{\rm RS}$
forces $K(\hat u,x)=K(\hat u,\hat x)$, hence $x=\hat x$ by (a).

Finally,
\begin{align*}
\Psi_\I^*(x)
=\sup_{c\ge0}\big\{(x-1)c+
\E\log\cosh(\sqrt{2c}\,z)\big\}.
\end{align*}
Combining this identity with \eqref{eq:appA-fixed-overlap} and
setting $2c=\alpha\beta^2r$ yields
\begin{align*}
\Phi(x)=\sup_{r\ge0}\mathcal A_{\alpha,\beta}(x,r),
\end{align*}
which proves \eqref{eq:AGS} with $q\in\mathcal D_\beta$.
For $x>0$, the maximizing $r$ is determined by
$\alpha\beta^2r=g^{-1}(x)$, while
\begin{align}
\Phi'(x)
=\tfrac12g^{-1}(x)-\frac{\alpha\beta^2x}{2D(x)^2}.
\label{eq:appA-AGS-derivative}
\end{align}
Thus interior critical points are exactly the admissible
spin-glass solutions of the AGS equations.
Conversely, any such critical point $x$ satisfies
$F_{u^*(x)}'(x)=0$ and $\partial_uK(u^*(x),x)=0$.
By (a), $x=x_{u^*(x)}$, and the envelope theorem gives
$\Lambda'(u^*(x))=0$. Uniqueness of the stationary radius then
forces $u^*(x)=\hat u$ and $x=\hat x$.
Therefore an admissible solution with $q>0$ exists if and only if
$\beta>\beta_g$, and it is unique. For $\beta\le\beta_g$,
$P_{\rm RS}=\Phi(0)=-\alpha\log(1-\beta)/2$.
The exchange proved in \eqref{eq:appA-minimax} concerns $u$ and $x$;
it does not exchange the $q$ and $r$ optimizations in \eqref{eq:AGS}.

\section{The replicon computation}\label{app:RSB}
We prove \eqref{eq:strict}, keeping the species weights explicit.
All path integrals below are over $v\in[0,1]$, and
$\la\pi,p\ra=\sum_{s=\I,\Sp}\int\pi^sp^s$.
Recall that
\begin{align}
\psi(\pi)&=\lambda_\I\psi_\I(\pi^\I)
                 +\lambda_\Sp\psi_\Sp(\pi^\Sp),\nonumber\\
\psi^*(p)&=\sup_\pi\{\la\pi,p\ra-\psi(\pi)\},\nonumber\\
f(t)&=\sup_p\left\{t\int p^\I p^\Sp-\psi^*(p)\right\}.
\label{eq:rb-functionals}
\end{align}
The paths $\pi^s$ are nonnegative and nondecreasing; the admissible
$p^s$ are nondecreasing with $0\le p^s\le\lambda_s$.

\emph{Trial bound and RS point.}
Convexity of $\psi$ in the paths gives, for $p=\partial_\pi\psi(\pi)$,
$\psi(\widetilde\pi)\ge\psi(\pi)+\la\widetilde\pi-\pi,p\ra$.
Consequently $\psi^*(p)=\la\pi,p\ra-\psi(\pi)$, and the admissibility
of this gradient yields
\begin{equation}\label{eq:trial}
f(t)\ge g_t(\pi):=\psi(\pi)-\la\pi,p\ra+t\int p^\I p^\Sp,
\qquad p=\partial_\pi\psi(\pi).
\end{equation}
At a constant path $c$, put $\Psi_s(c)=\psi_s(c\mathbf1)$.
The scalar derivatives are
\begin{align}
\Psi_\I'(c)&=\E\tanh^2(\sqrt{2c}\,z),\nonumber\\
\Psi_\Sp'(c)&=1-\bar b^{-1},\qquad
\bar b=\tfrac12(1+\sqrt{1+8c}).
\label{eq:rb-scalar}
\end{align}
Both range over $[0,1)$, so maximizing $g_t$ over constant pairs
gives $f_{\rm RS}(t)$, the restriction of \eqref{eq:rb-functionals}
to constant $p$. For $t>t_c$, its maximizer has $c_\I,c_\Sp>0$ and
\begin{equation}\label{eq:RScrit}
\begin{gathered}
c_\I=t\lambda_\Sp y,\qquad c_\Sp=t\lambda_\I x,\\
x=\Psi_\I'(c_\I),\qquad y=\Psi_\Sp'(c_\Sp).
\end{gathered}
\end{equation}
Thus $p_{\rm RS}=(\lambda_\I x,\lambda_\Sp y)$ and
$\pi_{\rm RS}=t(p_{\rm RS}^\Sp,p_{\rm RS}^\I)$ as paths.

\emph{One-step Hessians.}
Fix $0<\zeta<1$. Write $\rho=(\rho_1,\rho_2)$ for the path
equal to $\rho_1$ on $[0,\zeta)$ and $\rho_2$ on $[\zeta,1)$,
with $\Delta=\rho_2-\rho_1\ge0$.
For independent standard Gaussians $z,z'$,
\begin{equation}\label{eq:onestep}
\begin{gathered}
\psi_\I(\rho)=\rho_2-\zeta^{-1}\E\log\E'\cosh^\zeta X,\\
X=\sqrt{2\rho_1}\,z+\sqrt{2\Delta}\,z'.
\end{gathered}
\end{equation}
For the spherical species,
\begin{align}
\psi_\Sp(\rho)&=-\min_{b>2\zeta\Delta}D(\rho_1,\Delta;b),\nonumber\\
D&=\frac{\rho_1}{b-2\zeta\Delta}
 +\frac1{2\zeta}\log\frac b{b-2\zeta\Delta}\nonumber\\
&\hspace{7mm}+\frac12(b-1-\log b)-\rho_1-\Delta.
\label{eq:rb-spherical}
\end{align}
We expand at $\rho_1=\rho_2=c>0$, writing
$\rho_i=c+\varepsilon v_i$, $d=v_2-v_1$, and
$m=\zeta v_1+(1-\zeta)v_2=v_1+(1-\zeta)d$.
Derivatives at $\varepsilon=0$ are from the admissible side.

For Ising spins let $L=\log\cosh$, $T=\tanh$, and
$S=\operatorname{sech}^2$. The Gaussian heat identity
$\partial_a\E F(\sqrt{2a}\,z)=\E F''(\sqrt{2a}\,z)$ gives
\begin{align}
U_\Delta(h)&:=\zeta^{-1}\log\E'e^{\zeta L(h+\sqrt{2\Delta}z')},\nonumber\\
U_\Delta&=L+\Delta A_\zeta
       +\tfrac12\Delta^2B_\zeta+o(\Delta^2),\nonumber\\
A_\zeta&=L''+\zeta(L')^2=\zeta+(1-\zeta)S,\nonumber\\
B_\zeta&=A_\zeta''+2\zeta L'A_\zeta'\nonumber\\
 &=-2(1-\zeta)S^2+4(1-\zeta)^2T^2S.
\label{eq:rb-heat}
\end{align}
Applying the same identity to the outer expectation, and using
$L''''=-2S^2+4T^2S$ and $A_\zeta''=(1-\zeta)L''''$, yields
\begin{align}
\left.\frac{d^2}{d\varepsilon^2}\psi_\I(c+\varepsilon v)\right|_0
 &=-\E\{v_1^2L''''+2v_1dA_\zeta''+d^2B_\zeta\}\nonumber\\
 &=2\E S^2\{v_1^2+2(1-\zeta)v_1d+(1-\zeta)d^2\}
 \nonumber\\[-1mm]
 &\hspace{5mm}-4\E(T^2S)m^2.
\label{eq:rb-ising-expansion}
\end{align}
Here all functions inside expectations are evaluated at $\sqrt{2c}\,z$,
and the quadratic expression in braces is $\int v^2$.

For the sphere, $D_b(c,0;\bar b)=0$ gives
$\bar b(\bar b-1)=2c$. Along $(\rho_1,\Delta)=(c,0)+
\varepsilon(v_1,d)$, differentiation of $D_b=0$ gives
$b'=-D_{b\varepsilon}/D_{bb}$. The envelope theorem therefore gives
\begin{equation}\label{eq:rb-envelope}
\left.\frac{d^2}{d\varepsilon^2}\psi_\Sp(c+\varepsilon v)\right|_0
 =-D_{\varepsilon\varepsilon}
       +\frac{D_{b\varepsilon}^2}{D_{bb}}.
\end{equation}
The partial derivatives, evaluated at $(c,0;\bar b)$ while
holding $b$ fixed when differentiating in $\varepsilon$, are
\begin{align}
D_{\varepsilon\varepsilon}
 &=\frac{4\zeta v_1d+[2\zeta+4\zeta^2(\bar b-1)]d^2}{\bar b^2},
 \nonumber\\
D_{b\varepsilon}&=-\frac{v_1+[1+2\zeta(\bar b-1)]d}{\bar b^2},
 \qquad D_{bb}=\frac{2\bar b-1}{2\bar b^2}.
\label{eq:rb-envelope-derivatives}
\end{align}
Substitution in \eqref{eq:rb-envelope}, and
$\int v^2=v_1^2+2(1-\zeta)v_1d+(1-\zeta)d^2$, now give,
for both species,
\begin{equation}\label{eq:hess}
\left.\frac{d^2}{d\varepsilon^2}\psi_s(c+\varepsilon v)\right|_0
 =\kappa_s\int v^2-\varrho_s\left(\int v\right)^2,
\end{equation}
where
\begin{align}
\kappa_\I&=2\E\operatorname{sech}^4(\sqrt{2c}\,z),\nonumber\\
\varrho_\I&=4\E[\tanh^2(\sqrt{2c}\,z)
                         \operatorname{sech}^2(\sqrt{2c}\,z)],\nonumber\\
\kappa_\Sp&=\frac2{\bar b^2},\qquad
\varrho_\Sp=\frac{4(\bar b-1)}{\bar b^2(2\bar b-1)}.
\label{eq:rb-coefficients}
\end{align}
On the fixed one-step space the $L^2$ Hessian is consequently
$H_sv=\kappa_sv-\varrho_s(\int v)\mathbf1$.
In particular $H_sv=\kappa_sv$ when $\int v=0$.
The gradient $\partial_\pi\psi_s(\rho)$ in path space is itself one-step with the same break: by the cascade representation of the derivative \cite{CM25,CIM26,CM26}, $\la\kappa,\partial_\pi\psi_s(\rho)\ra$ is a Gibbs average of $\kappa(\alpha\wedge\alpha')$ times an overlap of two replicas, and under a one-level cascade the overlap depends on $\alpha\wedge\alpha'$ only through the cluster, $[0,\zeta)$ or $[\zeta,1)$. Its two values are
$\zeta^{-1}\partial_{\rho_1}\psi_s$ and
$(1-\zeta)^{-1}\partial_{\rho_2}\psi_s$, with the other level
held fixed. These weights are essential when using \eqref{eq:hess}.

\emph{An improving direction.}
Set
\begin{equation*}
w=(1-\zeta)\mathbf1_{[0,\zeta)}-\zeta\mathbf1_{[\zeta,1)},
\end{equation*}
so $\int w=0$ and $\int w^2=\zeta(1-\zeta)$, and set
$\pi_\varepsilon=\pi_{\rm RS}+\varepsilon(a_\I w,a_\Sp w)$.
Write $p_\varepsilon=\partial_\pi\psi(\pi_\varepsilon)$.
Differentiating \eqref{eq:trial}, the terms
$\la p_\varepsilon,\pi_\varepsilon'\ra$ cancel, leaving
\begin{equation*}
\frac{d}{d\varepsilon}g_t(\pi_\varepsilon)
 =\la-\pi_\varepsilon+t(p_\varepsilon^\Sp,p_\varepsilon^\I),
       p_\varepsilon'\ra.
\end{equation*}
The first factor vanishes at zero by \eqref{eq:RScrit}; hence
the first variation and the term containing $p_0''$ vanish.
With $K_s=\lambda_s\kappa_s$ evaluated at $c_s$, we have
$(p_0')^s=K_sa_sw$, and thus
\begin{align}
\left.\frac{d^2}{d\varepsilon^2}g_t(\pi_\varepsilon)\right|_0
 &=-\la\pi_0',p_0'\ra+2t\int(p_0')^\I(p_0')^\Sp\nonumber\\
 &=\zeta(1-\zeta)Q(a_\I,a_\Sp),\label{eq:d2g}\\
Q&=-K_\I a_\I^2-K_\Sp a_\Sp^2
                         +2tK_\I K_\Sp a_\I a_\Sp\nonumber\\
 &=-K_\Sp(a_\Sp-tK_\I a_\I)^2
           +K_\I(t^2K_\I K_\Sp-1)a_\I^2.
\label{eq:rb-square}
\end{align}
Only one positive direction is needed. If
$t^2K_\I K_\Sp>1$, choose $a_\I=-1$ and $a_\Sp=-tK_\I$:
the negative square is zero and $Q=K_\I(t^2K_\I K_\Sp-1)>0$.
Conversely, if $t^2K_\I K_\Sp\le1$, both terms in
\eqref{eq:rb-square} are nonpositive. This proves the criterion
\eqref{eq:replicon}. Since $w$ decreases and both $a_s<0$,
the chosen paths are nondecreasing and, for small
$\varepsilon>0$, nonnegative.

\emph{Strict instability.}
At the RS point, \eqref{eq:RScrit} and
$y=(\bar b-1)/\bar b$, $2c_\Sp=\bar b(\bar b-1)$ give
\begin{equation*}
t\lambda_\Sp=\frac{c_\I\bar b}{\bar b-1},\qquad
t\lambda_\I=\frac{\bar b(\bar b-1)}{2x}.
\end{equation*}
Consequently, with $\sigma^2=2c_\I>0$,
\begin{equation}\label{eq:rb-AT-ratio}
t^2K_\I K_\Sp=t^2\lambda_\I\lambda_\Sp\,\kappa_\I\kappa_\Sp
 =\frac{\sigma^2\E\operatorname{sech}^4(\sigma z)}
             {\E\tanh^2(\sigma z)}.
\end{equation}
To show that this ratio exceeds one, put $U=\sigma z$ and
$h(U)=\tanh^2U/U$, smoothly extended at zero. Gaussian
integration by parts gives
\begin{align*}
\E\tanh^2U&=\E[Uh(U)]=\sigma^2\E h'(U),\\
h'(U)&=\frac{2U\tanh U\operatorname{sech}^2U-\tanh^2U}{U^2}.
\end{align*}
Subtracting this identity from $\sigma^2\E\operatorname{sech}^4U$
completes another square:
\begin{equation}\label{eq:SOS}
\begin{split}
&\sigma^2\E\operatorname{sech}^4U-\E\tanh^2U\\
&\qquad=\sigma^2\E\left[
 \frac{(U\operatorname{sech}^2U-\tanh U)^2}{U^2}\right]>0.
\end{split}
\end{equation}
Indeed $F(U)=\tanh U-U\operatorname{sech}^2U$ is odd,
$F(0)=0$, and $F'(U)=2U\operatorname{sech}^2U\tanh U>0$
for $U>0$, so its only zero is zero. Also
$F(U)=\frac23U^3+O(U^5)$ and $\E U^4=3\sigma^4$,
so the deficit in \eqref{eq:SOS} is
$\frac43\sigma^6+O(\sigma^8)$.
Equations~\eqref{eq:d2g}--\eqref{eq:SOS} and the vanishing first
variation therefore imply
\begin{equation*}
f(t)\ge g_t(\pi_\varepsilon)
 =f_{\rm RS}(t)+\tfrac12\varepsilon^2\zeta(1-\zeta)Q
       +o(\varepsilon^2)>f_{\rm RS}(t)
\end{equation*}
for every $t>t_c$ and sufficiently small $\varepsilon>0$.

\emph{Optimization of the radius.}
Recall $u=s^2$, $t_u=(1+\alpha)\beta u/2$, and
$A(u)=\frac\alpha2[(\beta-1)u+1+\log u]$. Write
\begin{align*}
\Lambda_\alpha(u)&=A(u)-(1+\alpha)f(t_u),\\
\Lambda(u)&=A(u)-(1+\alpha)f_{\rm RS}(t_u).
\end{align*}
Both are strictly concave, since $f$ and $f_{\rm RS}$ are convex,
and $\Lambda_\alpha\le\Lambda$, strictly when $t_u>t_c$.
They attain their maxima: $\Lambda$ tends to $-\infty$ at both
ends by Appendix~\ref{app:LG}, and the same upper bound applies
to $\Lambda_\alpha$. For $\beta>\beta_g$, the unique maximizer
$\hat u$ of $\Lambda$ satisfies $t_{\hat u}>t_c$.
We then proceed by contradiction. If $P=P_{\rm RS}$, the maximizer $u^*$ of $\Lambda_\alpha$ would satisfy
$\Lambda(u^*)\ge\Lambda_\alpha(u^*)=P_{\rm RS}$,
forcing $u^*=\hat u$, where the inequality is strict.
Thus $P<P_{\rm RS}$.

To show also $t_{u^*}>t_c$, the boundary requires attention. Since $f=0$ on $[0,t_c]$ and $f$ is convex, $f'_+(t_c)\ge0$.
The mixed pressure is convex in its coupling amplitude $r$:
\begin{equation*}
G(r)=\varphi_\alpha(r^2)
 =\frac\alpha2r^2-(1+\alpha)
             f\left(\frac{1+\alpha}{2}r^2\right).
\end{equation*}
This follows from convexity of each finite-volume log partition
function in $r$ and passage to the limit. At
$r_c=\sqrt{2t_c/(1+\alpha)}$, convexity requires
\begin{equation*}
\alpha r_c=G'_-(r_c)\le G'_+(r_c)
 =\alpha r_c-(1+\alpha)^2r_c f'_+(t_c),
\end{equation*}
so $f'_+(t_c)=0$. Hence, with $u_c=(\beta\sqrt\alpha)^{-1}$,
\begin{equation*}
\Lambda'_{\alpha,+}(u_c)=A'(u_c)
 =\frac\alpha2[\beta(1+\sqrt\alpha)-1]>0.
\end{equation*}
Moreover $\Lambda_\alpha=A$ is increasing on $(0,u_c]$.
Thus $u^*>u_c$. At this radius $f(t_{u^*})>f_{\rm RS}(t_{u^*})$,
so no maximizing path pair in \eqref{eq:hopf} can be constant.
Finally, setting $f_{\rm 1RSB}(t)=\sup_{\pi\text{ one-step}}g_t(\pi)$
gives $f_{\rm RS}\le f_{\rm 1RSB}\le f$, with the first inequality
strict for $t>t_c$. The same radius argument therefore gives
$P\le P_{\rm 1RSB}<P_{\rm RS}$, as illustrated in Fig.~\ref{fig:gain}.


\begin{thebibliography}{35}
\bibitem{Hop82} J.~J.~Hopfield, \href{https://doi.org/10.1073/pnas.79.8.2554}{Proc.\ Natl.\ Acad.\ Sci.\ USA {\bf 79}, 2554 (1982)}.
\bibitem{AGS85} D.~J.~Amit, H.~Gutfreund, and H.~Sompolinsky, \href{https://doi.org/10.1103/PhysRevLett.55.1530}{Phys.\ Rev.\ Lett.\ {\bf 55}, 1530 (1985)}.
\bibitem{AGS87} D.~J.~Amit, H.~Gutfreund, and H.~Sompolinsky, \href{https://doi.org/10.1016/0003-4916(87)90092-3}{Ann.\ Phys.\ (N.Y.) {\bf 173}, 30 (1987)}.
\bibitem{PST94} L.~Pastur, M.~Shcherbina, and B.~Tirozzi, J.\ Stat.\ Phys.\ {\bf 74}, 1161 (1994).
\bibitem{BG98} A.~Bovier and V.~Gayrard, in \emph{Mathematical Aspects of Spin Glasses and Neural Networks}, Progr.\ Probab.\ {\bf 41} (Birkh\"auser, 1998), p.~3.
\bibitem{Tal98} M.~Talagrand, Probab.\ Theory Relat.\ Fields {\bf 110}, 177 (1998).
\bibitem{Tal11} M.~Talagrand, \emph{Mean Field Models for Spin Glasses}, Vol.~I (Springer, 2011).
\bibitem{BGG10} A.~Barra, G.~Genovese, and F.~Guerra, \href{https://doi.org/10.1007/s10955-010-0020-y}{J.\ Stat.\ Phys.\ {\bf 140}, 784 (2010)}.
\bibitem{BG08} A.~Barra and F.~Guerra, \href{https://doi.org/10.1063/1.3039083}{J.\ Math.\ Phys.\ {\bf 49}, 125217 (2008)}.
\bibitem{BGGT12} A.~Barra, G.~Genovese, F.~Guerra, and D.~Tantari, \href{https://doi.org/10.1088/1742-5468/2012/07/P07009}{J.\ Stat.\ Mech.\ (2012) P07009}.
\bibitem{BGG11} A.~Barra, G.~Genovese, and F.~Guerra, \href{https://doi.org/10.1088/1751-8113/44/24/245002}{J.\ Phys.\ A {\bf 44}, 245002 (2011)}.
\bibitem{BBSC12} A.~Barra, A.~Bernacchia, E.~Santucci, and P.~Contucci, \href{https://doi.org/10.1016/j.neunet.2012.06.003}{Neural Networks {\bf 34}, 1 (2012)}.
\bibitem{GT02} F.~Guerra and F.~L.~Toninelli, \href{https://doi.org/10.1007/s00220-002-0699-y}{Commun.\ Math.\ Phys.\ {\bf 230}, 71 (2002)}.
\bibitem{BCMT15} A.~Barra, P.~Contucci, E.~Mingione, and D.~Tantari, \href{https://doi.org/10.1007/s00023-014-0341-5}{Ann.\ Henri Poincar\'e {\bf 16}, 691 (2015)}.
\bibitem{Pan15} D.~Panchenko, \href{https://doi.org/10.1214/14-AOP967}{Ann.\ Probab.\ {\bf 43}, 3494 (2015)}.
\bibitem{CCM26} F.~Camilli, P.~Contucci, and E.~Mingione, Thermodynamic limit and variational formula for the Hopfield model with Gaussian patterns, forthcoming.
\bibitem{CM26} H.-B.~Chen and J.-C.~Mourrat, \href{https://arxiv.org/abs/2609.32710}{arXiv:2609.32710}.
\bibitem{CIM26} H.-B.~Chen, V.~Issa, and J.-C.~Mourrat, \href{https://arxiv.org/abs/2606.16636}{arXiv:2606.16636}.
\bibitem{GT15} G.~Genovese and D.~Tantari, \href{https://doi.org/10.1007/s11040-015-9181-x}{Math.\ Phys.\ Anal.\ Geom.\ {\bf 18}, 10 (2015)}.
\bibitem{GT20} G.~Genovese and D.~Tantari, \href{https://doi.org/10.1088/1751-8121/ab6b92}{J.\ Phys.\ A {\bf 53}, 094001 (2020)}.
\bibitem{CHI26} H.-B.~Chen, F.-H.~Ho, and V.~Issa, \href{https://arxiv.org/abs/2610.18744}{arXiv:2610.18744}.
\bibitem{BM26} J.~Boursier and J.-C.~Mourrat, \href{https://arxiv.org/abs/2610.18350}{arXiv:2610.18350}.
\bibitem{MouAHL} J.-C.~Mourrat, \href{https://doi.org/10.5802/ahl.77}{Ann.\ H.\ Lebesgue {\bf 4}, 453 (2021)}.
\bibitem{Mou21} J.-C.~Mourrat, \href{https://doi.org/10.2140/pmp.2021.2.281}{Probab.\ Math.\ Phys.\ {\bf 2}, 281 (2021)}.
\bibitem{Mou23} J.-C.~Mourrat, \href{https://doi.org/10.1214/22-AIHP1292}{Ann.\ Inst.\ Henri Poincar\'e Probab.\ Stat.\ {\bf 59}, 1143 (2023)}.
\bibitem{MouICM} J.-C.~Mourrat, Spin glasses and the Parisi formula, \href{https://arxiv.org/abs/2510.01054}{arXiv:2510.01054}.
\bibitem{CM25} H.-B.~Chen and J.-C.~Mourrat, Probab.\ Math.\ Phys.\ {\bf 6}, 1 (2025); \href{https://arxiv.org/abs/2311.08980}{arXiv:2311.08980}.
\bibitem{AT78} J.~R.~L.~de Almeida and D.~J.~Thouless, \href{https://doi.org/10.1088/0305-4470/11/5/028}{J.\ Phys.\ A {\bf 11}, 983 (1978)}.
\bibitem{Gue03} F.~Guerra, \href{https://doi.org/10.1007/s00220-002-0773-5}{Commun.\ Math.\ Phys.\ {\bf 233}, 1 (2003)}.
\bibitem{Ton02} F.~L.~Toninelli, \href{https://doi.org/10.1209/epl/i2002-00667-5}{Europhys.\ Lett.\ {\bf 60}, 764 (2002)}.
\bibitem{BGST18} A.~Barra, G.~Genovese, P.~Sollich, and D.~Tantari, \href{https://doi.org/10.1103/PhysRevE.97.022310}{Phys.\ Rev.\ E {\bf 97}, 022310 (2018)}.
\end{thebibliography}
\end{document}